\documentstyle[twoside,fleqn,espcrc2,fps]{article}

\newcommand{\AmS}{{\protect\the\textfont2
  A\kern-.1667em\lower.5ex\hbox{M}\kern-.125emS}}

\def\lsi{\raise0.3ex\hbox{$<$\kern-0.75em\raise-1.1ex\hbox{$\sim$}}}
\def\gsi{\raise0.3ex\hbox{$>$\kern-0.75em\raise-1.1ex\hbox{$\sim$}}}
\newcommand{\lsim}{\mathop{\lsi}}
\newcommand{\gsim}{\mathop{\gsi}}
\newcommand{\R}{{\kern+.25em\sf{R}\kern-.78em\sf{I} 
  \kern+.78em\kern-.25em}}

\title{Constructing Improved Overlap Fermions in QCD}

\author{W. Bietenholz
\address{ 
Institut f\"{u}r Physik, Humboldt Universit\"{a}t,
Invalidenstr. 110, D-10115 Berlin, Germany \\
$^{{\rm b}}$ Dept. of Physics, Univ. of Wuppertal, 
D-42097 Wuppertal, Germany \\
$^{{\rm c}}$ NIC, Forschungszentrum J\"{u}lich, D-52425 J\"{u}lich,
Germany }
, N. Eicker $^{{\rm b}}$, I. Hip $^{{\rm b,c}}$ 
and K. Schilling $^{{\rm b,c}}$
\thanks{Poster presented by W.B. at LATTICE 2000.}
}
\begin{document}

\begin{abstract}

We describe an explicit construction of approximate Ginsparg-Wilson
fermions for QCD. We use ingredients of perfect action origin,
and further elements. The spectrum of the lattice Dirac operator reveals 
the quality of the approximation. We focus on
$\beta =6$ for optimisation. Such fermions are intended to
be inserted into the overlap formula.
Hence we also test the speed of convergence 
under polynomial evaluation of the overlap formula.

\end{abstract}

\maketitle

\section{Introduction}
\vspace*{-1mm}

If a lattice Dirac operator $D$ obeys the Ginsparg-Wilson
relation (GWR) \cite{GW}
\begin{equation}
\{ D_{x,y},\gamma_{5} \} = 2 (D \gamma_{5} RD)_{x,y} \ ,
\end{equation} 
then it reproduces correctly the physical properties
related to chirality \cite{Has}. In particular, a GW
fermion has a chiral symmetry, which is lattice modified 
but exact \cite{ML}.

The kernel $R$ must be local, and $\{ R, \gamma_{5} \} \neq 0$.
For simplicity we assume 
\begin{equation}
\vspace*{-1mm}
R_{x,y} = \frac{1}{2\mu} \delta_{x,y} \quad (\mu >0),
{\rm ~and~} D^{\dagger} = \gamma_{5} D \gamma_{5}.
\vspace*{-1mm}
\end{equation}
Then the spectrum of $D$ is situated on a
circle in the complex plane: it is the circle through 0
with centre and radius $\mu$.

Classically perfect fermion actions solve the GWR \cite{Has}
and their scaling is excellent, but they are hard to construct.
They were applied successfully in the Schwinger model
\cite{LP,FL}, and the tedious work in $d=4$ is still in
progress \cite{Bern}.

There are, however, many more GW fermions:
starting from almost any sensible lattice Dirac operator $D_{0}$
(local, free of doublers etc.) we can generate a solution 
to the GWR by means of 
the {\em overlap formula}
\begin{equation} \label{overlap}
D_{ov} = \mu \Big( 1 + \frac{A}{\sqrt{A^{\dagger}A}} \Big) \ ,
\quad A = D_{0} - \mu \ .
\end{equation}
(The allowed range of $\mu >0$ will be commented on below).
If we start from the Wilson fermion, $D_{0} = D_{W}$, we arrive
at the Neuberger fermion \cite{Neu}. However, here
we are interested in improvements in various respects due to
a better choice of $D_{0}$ \cite{EPJC}.

Our guide-line for the search of a better $D_{0}$ is the following
observation:
if $D_{0}$ is a GW operator already (for some fixed GW kernel $R$),
then it coincides with the resulting overlap operator, 
$D_{ov} = D_{0}$.
In real life we are now going to construct an approximate GW operator
within a short range, and we use it as $D_{0}$. Hence the notorious
square root in eq.\ (\ref{overlap}) is approximately constant,
$\sqrt{A^{\dagger}A} \approx \mu$, and therefore
\begin{equation} \label{approx}
D_{ov} \approx D_{0} \ .
\end{equation}

For the free fermion, the perfect action can be constructed
and parametrised explicitly \cite{QuaGlu}, though it involves 
an infinite number of couplings. Still this fermion formulation
is local, because the absolute value of these couplings decays
exponentially with the distance between $\bar \psi$ and $\psi$.

We construct $D_{0}$ from the couplings of a truncated perfect
free fermion. The truncation is done so that only couplings inside
a unit hypercube survive, hence we arrive at a {\em hypercube fermion}
(HF). 
Then we gauge this HF (we distribute the coupling over the link
paths) by using just very few parameters,
in such a way that the violation of the GWR is minimised.
The criterion for this violation is proximity of the fermion 
spectrum to the unit GW circle, which corresponds to $\mu =1$.

Even if we only perform ``minimal gauging'' (shortest
lattice paths only, see below), the resulting operator $D_{HF}$ is 
promising for scaling and it is almost rotationally invariant 
\cite{BBCW} --- due to the perfect action background --- although 
it is short-ranged.


If we now insert $D_{0}=D_{HF}$ in the overlap formula, we expect 
a number of virtues for $D_{ov}$, which are all based on 
relation (\ref{approx}):

\vspace*{-2mm}

\begin{itemize}

\item $D_{ov}$ is promising for a good {\em scaling behaviour} 
and {\em approximate rotational invariance}, properties which
are inherited from $D_{0}$.

\item We expect $D_{ov}$ to have a high level of locality
(a fast exponential decay of the correlations), since
$D_{0}$ is ultralocal and long-range couplings can be turned 
on just slightly by the overlap formula.

\item The iterative transition from $D_{0}$ to $D_{ov}$
is fast, since we already start off in the right vicinity.

\end{itemize}

\vspace*{-2mm}

This is in contrast to the Neuberger fermion, which does
a rather poor job with respect to all these properties.
The above predictions have been tested and confirmed
in a comprehensive study of the 2-flavour Schwinger model
\cite{BH}. We used a 2d HF, which had (amazingly) a similar
scaling quality as the very mildly truncated, classically
perfect fermion of Ref.\ \cite{LP}.
The superiority of the resulting $D_{ov}$ (improved
overlap fermion) over the Neuberger fermion (or standard
overlap fermion) was striking in all the respects listed
above.

We are now carrying on this program to (quenched) QCD 
\cite{prep}. Locality properties of free
4d improved overlap fermions have already been discussed
in Ref.\ \cite{EPJC}, and first results for QCD
have been reported in Ref.\ \cite{China}. The same concept 
has also been adopted in Ref.\ \cite{TDG}, which presents
some results with respect to the last property (speed of the
transition), based on a very simple choice of $D_{0}$.
On the other hand, sophisticated and complicated
approximate GW fermions have been constructed in Ref.\ \cite{Gatt}.

Following the same lines, one may also work on improved domain wall
fermions by optimising the choice of the 4d operator that corresponds
to $D_{0}$ \cite{EPJC}. Recent proposals focus on accelerating
the transition to exact chirality, which is expressed here in terms
of the required extent of the extra dimension \cite{DWF}.

\vspace*{-1mm}

\section{Construction of $D_{0}$}

We now describe explicitly the construction of our HF, which 
approximates a GW fermion, and which is also promising in other 
respects. We proceed gradually in a sequence of steps.

\subsection{Truncated perfect free fermion}

As we mentioned earlier, perfect actions for free fermions
can be constructed analytically, where the term $R$ occurs
in the renormalisation group transformation term.
The locality is optimal for $R_{x,y} = \frac{1}{2}\delta_{x,y}$
(i.e. $\mu =1$) \cite{QuaGlu}, which is the standard GW kernel,
and which we are going to use in the following. 

We write the Dirac operator of the free perfect fermion as
$D(x-y) = \rho_{\mu}(x-y)\gamma_{\mu}+\lambda (x-y)$, where
we denote $\rho_{\mu}$, $\lambda$ as the vector term and the scalar
term, respectively. We truncate both terms by imposing periodic
boundary conditions, so that we arrive at a free HF.
The explicit couplings
inside a unit hypercube are given in Ref.\ \cite{BBCW}, Table 1.
The scaling of this free HF is excellent \cite{BBCW}, and its
spectrum is very close to a GW unit circle \cite{EPJC}, hence
it is a good approximation to a GW fermion.

The resulting free overlap HF has the expected virtues,
in particular good scaling and locality \cite{EPJC}. \\

\vspace*{-2mm}

\subsection{Minimal gauging}

We now proceed to quenched QCD, and we always use the plaquette
gauge action. Tests with improved gauge actions are on the way.
\footnote{They seem to be indeed useful for chirality, see
last reference in \cite{Gatt}.}

Our point of departure is a {\em minimal gauging}
of the HF: the free couplings are attached to the shortest
lattice paths only, in equal parts where several shortest
paths coexist. This is the simplest version of a HF in gauge 
theory, but the harmony of the truncated free couplings
leads to a beautiful pion dispersion relation, even at
strong coupling, see Ref.\ \cite{BBCW}, Fig.\ 10.
On the other hand, this kind of gauging implies a strong
mass renormalisation --- comparable to the Wilson fermion.
The reason is that all paths are suppressed by the gauge field
(compared to the free case), whereas the coupling between 
$\bar \psi$ and $\psi$ on the same site in unaltered.
For instance, at $\beta=5$ we obtain a ``pion mass'' of 
$3.0$ \cite{BBCW}.

\begin{figure}[hbt]
\vspace*{-10mm}
\def\fpsangle{270}
\epsfxsize=53mm
\fpsbox{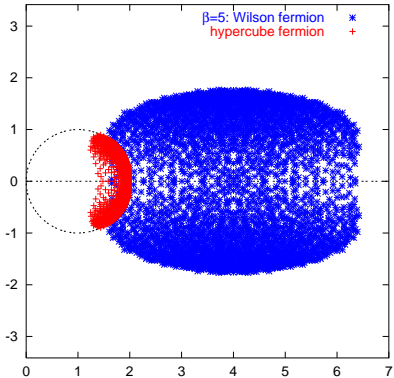}
\vspace{-10mm}
\caption{\it{The fermionic spectrum for the minimally gauged HF
and for the Wilson fermion, in a typical configuration at
$\beta =5$.}}
\label{Fb5}
\vspace*{-8mm}
\end{figure}

This effect is also visible from the fermionic spectrum:
again at $\beta =5$ the right arc still follows closely 
the GW circle --- unlike the Wilson fermion spectrum,
which extends to large real parts. However, the physically 
crucial left arc is absent in both cases, see Fig.\ \ref{Fb5}. 
\footnote{Note that this is a situation where it could be 
misguiding to consider only the spectrum of $A^{\dagger}A$, 
as it is sometimes done in the literature.}

In such a situation, the overlap formula --- which is supposed to
provide a projection of the spectrum onto a GW circle --- is not
applicable any more: there is no suitable value of $\mu$, which
is the centre of the circle to be mapped on. Wherever we choose 
it, we would be confronted frequently with too many mappings to 
the left (revival of the doubling problem) or to the right arc
(mass renormalisation is back) or both.

Now it is of interest where the applicability of the overlap 
formula sets in. 
At $\beta \approx 5.4$ the coupling
is still too strong, whereas at $\beta \gsim 5.6$ we are statistically
on safe grounds for a simple operator $D_{0}$ \cite{China}.
\footnote{If we deform a configuration continuously into another
topological sector, then at least one eigenvalue has to cross
the centre of the circle, since we define the topological charge by
the index theorem. What we achieve at $\beta \gsim 5.6$ is that
this transition is very quick, i.e.\ for most configurations occurring
at that coupling strength, all (almost) real eigenvalues clearly
belong to the left or the right arc.}

So far, we show the full fermionic spectra on small $4^{4}$
lattices. From our experience, this is sufficient to obtain
the qualitatively correct impression about the performance of
a lattice Dirac operator. The fact that the arc close to zero 
tends to be absent is {\em not} due to the fermion formulation,
but solely due to the small lattice. Later we will show
this arc specifically also on $8^{4}$ lattices, in order to
provide a more complete picture.

\subsection{Criticality through link amplification}

We now go beyond minimal gauging by introducing a few
extra parameters and optimising them so that the GWR
violation is minimised at $\beta =6$.

\begin{figure}[hbt]
\vspace*{-9mm}
\def\fpsangle{270}
\epsfxsize=53mm
\fpsbox{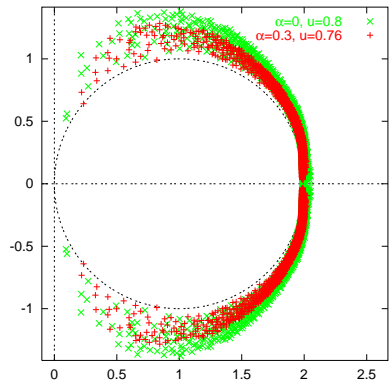}
\vspace*{-10mm}
\caption{\it{HF spectra at $\beta=6$ on a $4^{4}$ lattice at
critical link amplification $1/u$ without fat links ($\alpha =0$)
and with fat links ($\alpha = 0.3$).}}
\label{Fucrit}
\vspace*{-10mm}
\end{figure}

First we attach a {\em amplification factor} $1/u$ ($u \lsim 1$)
to {\em each link}. The idea is to compensate the mean link
suppression due to the gauge field. 
At $\beta =6$ we obtain 
criticality at $u \simeq 0.8$. This already provides a decent
approximation to a GW fermion, see Fig.\ \ref{Fucrit}.
\vspace*{-2mm}

\vspace*{-2mm}

\subsection{Fat links and vector term suppression}

We now introduce {\em fat links} by globally substituting all 
links as
\begin{displaymath}
link \rightarrow (1-\alpha ) \, link + \frac{\alpha}{6} 
\Big[ \sum staples \Big] \ .
\end{displaymath}
It turns out that for instance $\alpha = 0.3$ is a useful value:
it helps to move the upper (resp.\ lower) arc closer to the circle,
and it also pulls the eigenvalues closer together, which are both
desired effects, as illustrated in Fig.\ \ref{Fucrit}.

On the other hand, a sequence of experiments suggests that
the {\em clover term} is not a very powerful tool to improve 
the GWR approximation. Hence we omit it, since we only want to 
include new ingredients if they really yield a significant 
progress.

\begin{figure}[hbt]
\vspace*{-8mm}
\def\fpsangle{270}
\epsfxsize=60mm
\fpsbox{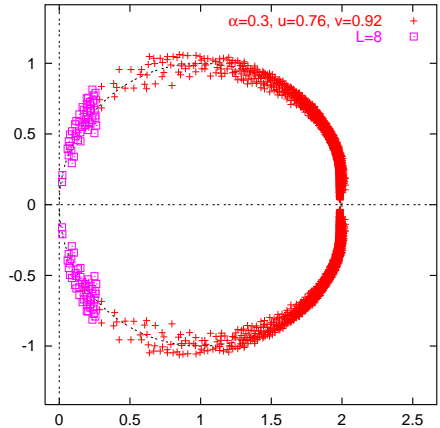}
\vspace{-10mm}
\caption{\it{The HF on a $4^{4}$ lattice with critical link
amplification, fat links and a suppression of the 
vector term. We also show the ``continuation'' around $0$
on a $8^{4}$ lattice with the same parameters, i.e.
$u=0.76$, $\alpha =0.3$, $v=0.92$.}}
\label{Fopt}
\end{figure}

An exception from that rule are the factors that we attach to
the links, since they require practically no computational effort.
Hence we can also introduce different link factors for the
scalar term $\lambda$ and the vector term $\rho_{\mu}$.
The former is responsible for the chiral limit, so we
leave it unaltered at $u=0.8$. However, we suppress
the vector term a little, because it is responsible for
the imaginary part of the fermionic spectra, which are a
bit too large so far. So now we multiply in $\rho_{\mu}$ only
the links by $v/u$, $v \lsim 1$. By tuning the value of $v$
we can make the above spectra follow the shape of the GW circle.
Still the fat link is useful to suppress the radial fluctuations
of the eigenvalues. We use again $\alpha =0.3$, and find the
suitable vector link suppression to be $v = 0.92$.
This leads to the quite satisfactory spectrum shown in 
Fig.\ \ref{Fopt}.

\section{Convergence to $D_{ov}$}
\vspace*{-1mm}

In $d=4$ the troublesome square root in the overlap formula
(\ref{overlap}) can only be approximated iteratively.
A simple approach is the expansion of the inverse
square root in a series in $A^{\dagger}A - \mu^{2}$.
This converges for all eigenvalues only if we start from
a good GW fermion approximation $D_{0}$, and then the
convergence is fast, as we have shown in the 
Schwinger model \cite{BH}.

We now turn our attention to the usual method
which approximates the sign function $\epsilon$ in
\vspace*{-1mm}
\begin{equation} \vspace*{-1mm} \label{sign}
D_{ov} = \mu [ 1 + \gamma_{5} \epsilon (H)] \ , \
H = \gamma_{5}(D_{0}-\mu ) = H^{\dagger}
\vspace*{-1mm}
\end{equation}
\vspace*{-1mm}
by a polynomial in $H$. 

\begin{figure}[hbt]
\vspace*{-9mm}
\def\fpsangle{0}
\epsfxsize=53mm
\fpsbox{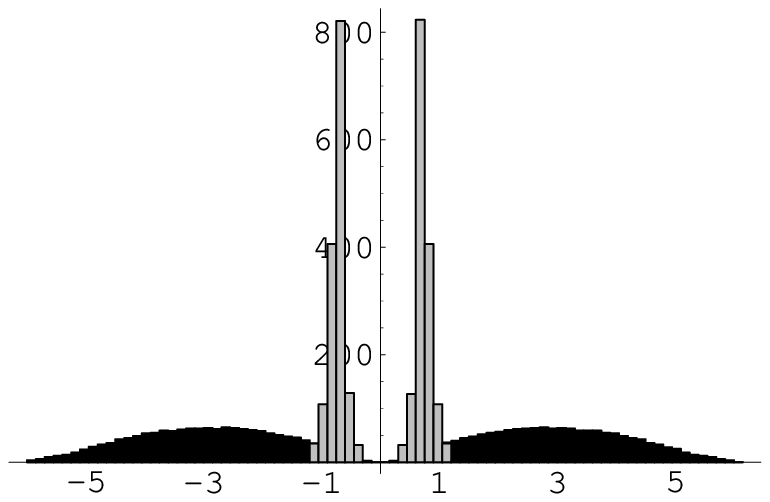}
\def\fpsangle{0}
\epsfxsize=53mm
\fpsbox{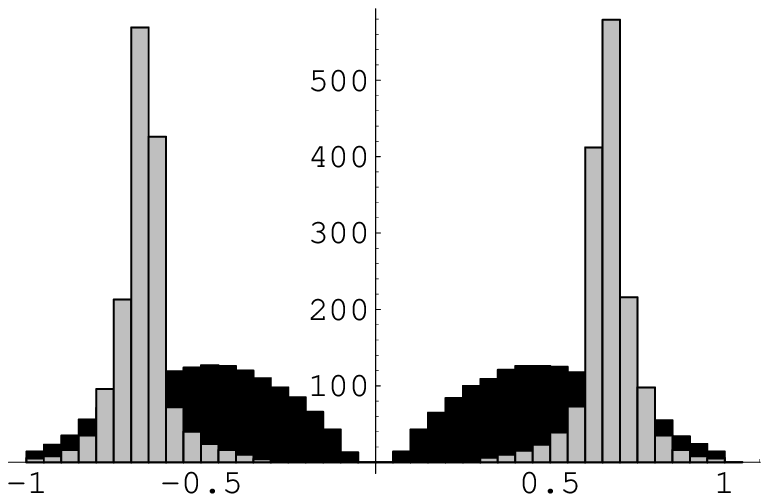}
\vspace*{-10mm}
\caption{\it{On top: The eigenvalue histograms for typical 
spectra of $H_{HF}$ (grey) and $H_{W}$ (black) at $\beta =6$. 
Below: The same after
re-scaling so that all $\vert$eigenvalues\,$\vert$ are $\leq 1$.}}
\label{histo}
\vspace*{-10mm}
\end{figure}

The histograms for typical spectra of $H_{W}$ and
$H_{HF}$ at $\beta =6$ are shown in Fig.\ \ref{histo} (on top). 
For the HF (with $\alpha =0.3$, $u=0.76$, $v=0.92$) 
the spectrum is already nicely
peaked near $\pm 1$, whereas the Wilson fermion yields a
broad distribution.

\begin{figure}[hbt]
\def\fpsangle{270}
\epsfxsize=55mm
\fpsbox{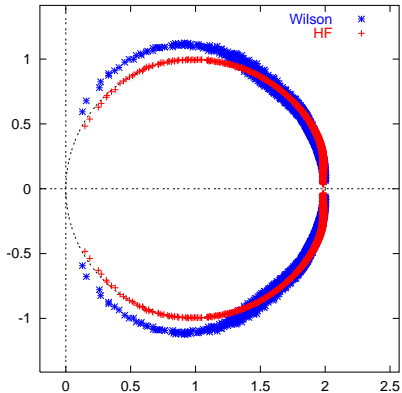}
\vspace{-10mm}
\caption{\it{The spectra of approximate overlap fermions, 
where the sign function is replaced 
by a polynomial of degree 21. We use a
configuration typical at $\beta =6$ and start from 
$D_{0}=D_{W}$ (stars) resp.\ from $D_{0}=D_{HF}$ (crosses).}}
\label{4dpoly}
\vspace*{-8mm}
\end{figure}


In order to make the polynomial approximation
applicable for all eigenvalues,
we have to re-scale the spectra such that they are confined
in the interval $[-1,1]$, see Fig.\ \ref{histo} (below). 
This does not affect the spectrum
of $H_{HF}$ too much, and in particular there is a large gap 
around 0, where any polynomial approximation of the sign
function has its largest error.
On the other hand, for $H_{W}$ there is a significant
eigenvalue density in the vicinity of 0, which shows again
that here it is far more demanding to enforce convergence
to a GW fermion.

We illustrate the difference by using a
linear combination of Chebyshev polynomials (as suggested
in Ref.\ \cite{HJL}) with maximal degree 21 
to approximate the sign function in the overlap
formula (\ref{sign}).
\footnote{A restriction to degrees below $\approx 20$
is not sensible, regardless what $D_{0}$ one uses,
because the approximation to the sign function is
too poor. This is also true for other polynomials
that may be used for this purpose.}
For a typical configurations at $\beta =6$ 
(on a $4^{4}$ lattice) this leads to the spectra
shown in Fig.\ \ref{4dpoly} for the Wilson fermion resp.\ the HF.
We see that the latter is clearly superior.
This gain might already compensate
the computational overhead by a factor of $O(10)$ in the 
application of $D_{HF}$.


\vspace*{-3mm}

\section{Conclusions}

\vspace*{-1mm}

A good approximation to a GW fermion in QCD at $\beta =6$
has been constructed, using only 10 independent parameters.
This HF is ready to be inserted in the overlap formula
(\ref{overlap}). This works down to $\beta \approx 5.6$,
and it leads to an exact GW fermion. There are good reasons
to expect that the resulting GW fermion is superior over the
standard Neuberger fermion with respect to scaling, rotational
symmetry and locality. We tested the speed of 
convergence towards an overlap fermion if the sign function
in the overlap formula is replaced by Chebyshev polynomials.
Also with this respect we found a significant gain if we start 
from $D_{HF}$ instead of $D_{W}$. This effect
will be discussed in detail in Ref.\ \cite{prep}.

\end{document}